\documentclass[11pt]{jfm}
\usepackage{multicol}
\usepackage[english]{babel}
\usepackage{amsfonts}
\usepackage{graphicx}
\usepackage{rotating}
\usepackage[a4paper]{geometry}
\usepackage{multirow}
\usepackage[table]{xcolor}
\usepackage{float}
\usepackage{amsmath}
\usepackage{wrapfig}
\usepackage{hyperref}
\usepackage{mathtools}
\usepackage{amssymb}
\usepackage{amsmath}
\usepackage{amsbsy}
\usepackage[font=footnotesize]{caption}
\usepackage{textcomp}
\usepackage{natbib}
\usepackage{breqn}
\usepackage{subfig}
\usepackage{physics}
\usepackage{tikz}
\usetikzlibrary{calc}

\def\BV{Brunt-V{\"a}is{\"a}l{\"a}\ }

\DeclareGraphicsExtensions{.pdf,.png}
\title{
Particle transport induced by internal wave beam streaming in lateral boundary layers}
\author{E. Horne\aff{1,4}\corresp{\email{ernesto.horne@ladhyx.polytechnique.fr}}, F. Beckebanze\aff{2}\corresp{\email{f.beckebanze@uu.nl}}, D. Micard\aff{3,4}, P. Odier\aff{4}, L. R. M. Maas\aff{5}, S. Joubaud\aff{4}}
\affiliation{
\aff{1}LadHyX, CNRS, \'{E}cole Polytechnique, 91128 Palaiseu CEDEX, France \\
\aff{2}Mathematical Institute, Utrecht University, P.O. Box 80010, 3508 TA Utrecht, The Netherlands \\
\aff{3}LMFA UMR 5509 CNRS, Universit\'{e} de Lyon, \'{E}cole Centrale 69130 \'{E}cully Lyon, France
\aff{4}Univ Lyon, ENS de Lyon, Univ Claude Bernard, CNRS, Laboratoire de Physique,F-69342 Lyon, France
\aff{5}Institute for Marine and Atmospheric research Utrecht (IMAU), Utrecht University, Princetonplein 5, 3584 CC Utrecht, The Netherlands}
\date{March 2017}
\begin{document}
\maketitle

\begin{abstract}
Quantifying the physical mechanisms responsible for the transport of sediments, nutrients and pollutants in the abyssal sea is a long-standing problem, with internal waves regularly invoked as the relevant mechanism for particle advection near the sea bottom. This study focuses on internal-wave induced particle transport in the vicinity of (almost) vertical walls. We report a series of laboratory experiments revealing that particles sinking slowly through a monochromatic internal wave beam experience significant horizontal advection. Extending the theoretical analysis by \cite{Be18}, we attribute the observed particle advection to a peculiar and previously unrecognized streaming mechanism originating at the lateral walls. This vertical boundary-layer streaming mechanism is most efficient for strongly inclined wave beams, when vertical and horizontal velocity components are of comparable magnitude. We find good agreement between our theoretical prediction and experimental results.

\end{abstract}


\section{Introduction}

Internal waves are ubiquitous in the global oceans where they play a critical role in transporting sediments, nutrients and pollutants from localized sources to remote places \citep{Al03}, with potentially strong influence on marine ecosystems \citep{Wo18}. Many efforts have been undertaken to understand the behavior of particles in the ocean. For the upper ocean, particle advection, dominated by ocean currents and surface waves, is reasonably well-understood \citep{Se18}. 
On the contrary, the mechanisms dominating suspended particle advection in the stratified interior and near the bottom of the abyssal ocean remain to be established clearly. 
In analogy to surface waves being important for particle advection near the surface, it is regularly invoked that internal waves must be relevant for particle advection near the sea bottom, both vertically through mixing related to breaking of waves and horizontally through induced (Lagrangian) transport.
While internal-wave induced sediment transport has been observed on continental shelves \citep{HosegoodGRL04, Butman06, QuaresmaMG07}, field observations in the abyssal sea are largely hampered by the technical challenges. 
\\
\indent 
Recent studies indicate that mixing in the vicinity of steep ocean topography is much stronger than previously thought  \citep{MF17, M17}, suggesting strong erosion of nearly neutrally buoyant particles into the water column right above the bottom.  Coincidentally, internal wave motion is typically also enhanced above rough topography \citep{WF04, GK07}. This raises the question of how internal wave motion may facilitate advection of slowly sinking particles in the vicinity of steep topography. This study investigates boundary layer effects near steep topography on internal-wave induced particle advection.
Steep refers to the slope $\tan \alpha$ of the topography with respect to the horizontal being significantly larger than the slope $\tan \theta$ of the internal wave propagation. 
\\
\indent  
We analyse a series of laboratory experiments (partially reported in \textit{Ch.} 5 in \cite{Ho15}) revealing that slowly sinking
particles experience significant horizontal advection in the vicinity of  
an internal wave beam. Peculiarly, the observed particle advection is strongly dependent on the wave frequency $\omega_0$ relative to the \BV frequency, $N_0=\sqrt{-\frac{g}{\langle \bar{\rho} \rangle}\dv{\bar{\rho}}{z}}$, where $g$ is the acceleration of gravity, $\bar{\rho}(z)$ is the component of the density that monotonically increases with depth and is stationary, $\langle \bar{\rho} \rangle$ is the vertical average of $\bar{\rho}(z)$; therefore, the full density profile can be written as $\rho_0=\bar{\rho}(z)+\rho$, where $\rho$ represents the density perturbations.  The goal of this study is to understand and rigorously describe the dynamics that dominates the experiments. We provide strong evidence that the observed particle displacement is facilitated by an internal-wave induced horizontal mean flow. 
 \\
\indent  
Our theoretical analysis, extending the work by \cite{Be18} to weakly non-linear internal waves, highlights a previously unrecognized lateral-wall streaming mechanism. 
Streaming refers to irreversible mean flow generation through non-linear internal wave interactions, in analogy to acoustic streaming \citep{Li78}. Typically, streaming results in \emph{strong} horizontal mean flow generation if mean vertical vorticity is produced. Here, `strong' refers to persistent, cumulative transfer of energy from the wave field into the mean flow. Strong mean flow generation is known to occur due to horizontal cross-beam variation \citep{Bo12, KA15, Se16, Be18b}, with important modifications by planetary rotation \citep{GB12, FKA18}, and upon reflection where incident and reflected beams interact \citep{Thorpe97, Gr13, ZD15, Raja18}. \cite{RV18} recently also found strong mean flow generation in a flat bottom boundary layer. 
\\
 \indent 
 Our detailed analysis demonstrates that the lateral-wall streaming is related to a peculiar difference in lateral boundary layer \emph{thickness} for vertical and horizontal velocity components, first noted by \cite{VC03}, and recently linked to intensified wave field damping by \cite{Be18}. The lateral-wall streaming is strongest for substantially inclined wave beams, when vertical and horizontal along-wall velocity components are of similar magnitude, explaining the experimentally observed dependency on the wave frequency.  We remark that the previously unknown lateral-wall streaming was only recognized due to the \emph{absence} of known streaming mechanisms in the experimental set-up. As such, it is likely to have also occurred in other laboratory experiments on internal waves, among which the experiments on internal wave attractors by \cite{Ha10} and \cite{Br17}. 
\\
\indent 
The structure of this paper is as follows. In \S \ref{exp} we report the laboratory set-up, including a detailed description of a newly developed particle injector in \S\ref{setup_particles}. The experimental results (\S \ref{obs}) are the motivation for the theoretical derivation of the internal-wave induced mean flow near the lateral walls, presented in \S \ref{theory}. The multiple scale analysis may be skipped as the main theoretical results are summarized in \S \ref{comparison}, where we compare them to the experimental results. Oceanic circumstances for which our results are potentially important, as well as possible extensions and limitations of our study are discussed in \S \ref{con}.


\begin{figure}
\begin{center}
{\includegraphics[width=0.9\textwidth]{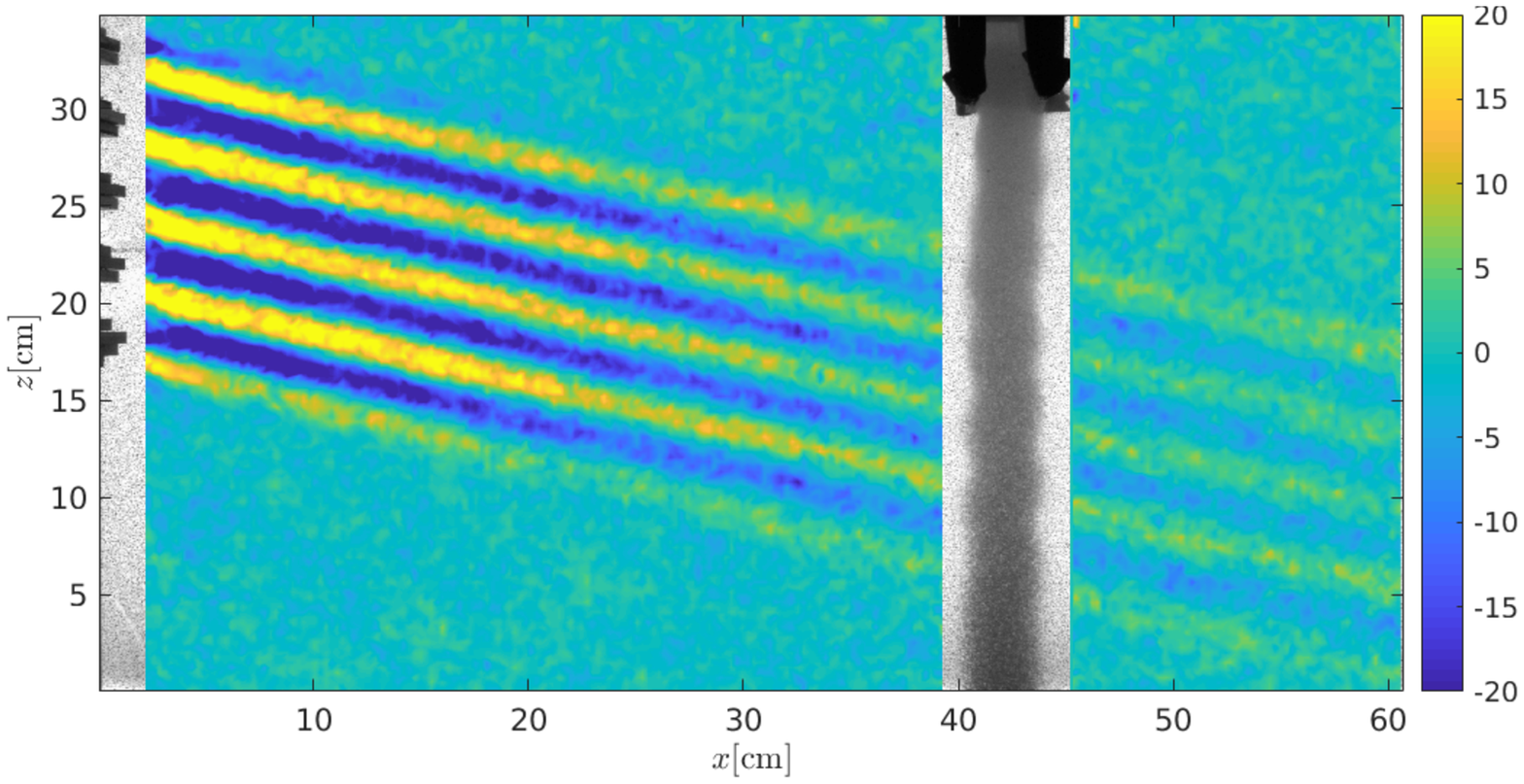}\put(-50,192){$ \partial_x \rho $ [kg/m$^4$]}
\put(-160,192){\scriptsize particle injector}
\put(-244, 68){\scriptsize $\theta$} 
\put(-243, 66){\line(1,0){29}}
\put(-245, 75){\rotatebox{-15}{ \line(1,0){29} }}
\put(-375, 179){(a)}
}
{\hspace*{-0.48cm} \includegraphics[ width=0.97\textwidth]{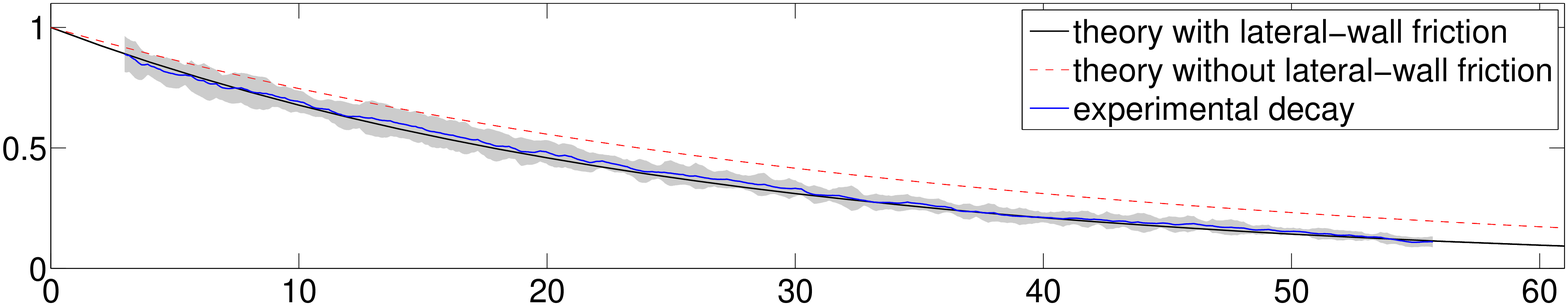}
\put(-36,11){\tiny $\leftarrow$ 0.09}
\put(-36,16){\tiny $\leftarrow$ 0.16}
\put(-205,-6){\scriptsize $x$ [cm]}
\put(-380, 55){(b)}
}\hspace{0.39cm}
\caption{\small{(a): Snapshot of experimental internal wave beam, generated by wave maker on the left ($x=0$~cm), propagating downward to the right with an angle $\theta=\sin^{-1}[\omega_0/N_0]=15^{\circ}$ with respect to the horizontal, and intersecting the sinking particle column, here at $x=40$ cm from the wave maker. 
The color map visualizes the horizontal density gradient, $\partial_x \rho$ (here, the vertical background density gradient is $ 72 $ kg/m$^4$). (b): Spatially averaged along-beam decay of the wave field (normalized) extracted from $\omega_0$-filtered time series (blue), with the gray shading indicating two standard deviations. The observation matches the theoretical exponential decay (derived in \S \ref{theory}) including both internal shear and lateral wall-friction (black line). Taking only internal shear dissipation into account (red dashed line) over-predicts the wave beam strength at $x=60$ cm by about $80$\% ($\approx 1-0.16/0.09$).    }}
\label{fig:f1}
\end{center}
\end{figure}

\section{Experimental set-up}
\label{exp}
A rectangular tank of inner size $L \times W \times H= 156\times 17\times 42.5 ~$cm$^{3}$, corresponding respectively to coordinates $x$, $y$ and $z$, is filled up to $\sim$37 cm with a linearly salt-stratified fluid by using the two-bucket method~\citep{Fortuin1960, Oster1963}. Vertical density profile measurements are performed with a conductivity probe. The fluid densities at $z=33.6$ cm, where the particles enter the fluid, and bottom ($z=0$) are $\rho_0 = 0.998$ g/cm$^{3}$ and $1.039 \pm 0.002$ g/cm$^3$ respectively, corresponding to buoyancy frequency $N_0 = 1.1 \pm  0.03$ rad/s. 
Internal plane waves are created by an internal wave generator based on the set-up developed by \cite{Go07}. The wave generator consists of 50 plates of height $6.5$ mm, of which we use the upper $25$ plates (from $z=17.35$ cm to $z=33.6$ cm) for the forcing in the present experiments. We install the plates such that the oscillating plates mimic a sinusoidal vertical profile, with upward phase propagation and vertical wave length $L_z=3.9$~cm composed by $6$ plates, over the vertical extent of $4.17 L_z=16.25$~cm. 
Fig. \ref{fig:f1}a shows a composite snapshot of the steady state wave beam and the column of settling particles described in \S \ref{setup_particles}, here at $x=40$ cm. The colormap visualizes the horizontal density perturbation, $\partial_x \rho$, derived from Synthetic Schlieren, which can be converted to any wave beam field quantity. 

Eleven experiments are performed for imposed wave frequency $\omega_0$ in the range $0.125 - 0.785$ rad/s, all for fixed wave maker amplitude $A_0 = 0.9$ cm. The generated internal wave beam with characteristic wave length $L_0=L_z \cos \theta$ propagates downwards and to the right at angles $\theta=\sin^{-1}[\omega_0/N_0] = 0.11 - 0.8$ rad with respect to the horizontal. Table \ref{t1} summarizes the parameter values and ranges explored in this work. 
The along-beam decay (Fig. \ref{fig:f1}b) is extracted from a time series of 175 s (frames at 4 Hz), starting 50 s after the on-set of the wave maker, filtered at the forcing frequency $\omega_0=0.22$ rad/s for this case, and spatially averaging over 2 cross-beam wave lengths (300 data points). The exponential decay only matches the theoretical decay rate upon incorporating lateral-wall friction, emphasizing the importance of the lateral walls in the present experiments. The theoretical decay rates are derived in \S \ref{theory}.
Viscous attenuation reduces the wave beam strength by $\sim 99.7$\% at the end of the tank ($x=156$ cm), making a sponge layer unnecessary. 
The wave field is observed using Synthetic Schlieren technique, with a computer-controlled video camera (Allied Vision
Technologies Stingray) with CCD matrix of $2452$ $\times$ $2054$ pixels, which measures the variations of the gradient of density of the fluid~\citep{Dalziel2000, Sutherland1999}. The conversion of pixels to distance varies by about $5$\% throughout the tank, being $26$ pix/cm at the back ($y = -W/2$) and $28.5$ pix/cm at the front ($y=W/2$) of the tank. We apply the conversion $26.8$ pix/cm to the experimental data, to match experimentally determined wave beam height (in pixel) to the theoretically known wave beam height.

\subsection{\label{setup_particles}Sinking particle column}
 We developed a particle injector that allows for the production of a column of particles of controlled particle density, raining into the stratified fluid in a rectangular section, which embraces the full width of the tank (from wall to wall). The particle injector is placed at $18$, $25$ or $40$ cm to the right of the wave generator. It is only at $18$ cm that the particle column is located outside of the wave beam bottom reflection region for all tested wave frequencies, in which case we can rule out streaming effects associated with interaction of incident and reflected beams. For this reason, we only analyze experimental particle column dynamics injected at 18 cm distance from the wave maker.  \\
 \indent 
The particles are injected between two vertical acrylic plates (see Fig. 1a) and slightly below the free surface through a slit in a copper tube,  the slit extending along the entire $y$-direction, from wall to wall. The distance between the acrylic plates (in along wall $x$-direction) is fixed at 3 cm for the present experiments. Beforehand, the particles are mixed with fresh water (density $\rho_0 = 0.998$ g/cm$^{3}$ ) and surfactant to avoid clustering during the immersion in the stratified fluid. The fresh water, acting as a carrier for the particles, drags the suspended particles from a small container (particle reservoir) through the copper tube to its slit transect. A peristaltic pump guarantees that the same amount of fluid is pumped into the slit transect of the copper tube, as is sucked out of it. 
A constant flow rate of $20$~mL/min for the carrier fluid results - after an initial transient phase - in a stationary homogeneous column with packing fraction $\phi \sim \mathcal{O}(10^{-2}$). We verified experimentally that this small packing fraction does not affect the internal wave field.  
\\
\indent 
The granular column consists of polystyrene grains with density $\rho_p = 1.055\pm 0.01$ g/cm$^3$\footnote{The particle density reported by the particle producer and in \cite{Ho15} is $1.05$ g/cm$^3$, without any margin of error. Our own measurements - inferring particle densities from the height where particles come to rest in a linearly stratified fluid  - revealed a slightly larger average particle density, with an estimated error of $0.01$ g/cm$^3$.}. The particle size of $\mathcal{O}(100)$ $\mu$m was intended to be identical in all experiments. Indirect measurements of the prevailing particle sizes, derived from an experimental Doppler Shift and explained in the appendix, \S \ref{particle_trajectory}, indicate that the prevailing particle diameter increased in consecutive experiments, ranging from $40$ to $240$ $\mu$m. We believe that the carrier fluid segregated the particles, dragging smaller particles more efficiently, thereby first removing predominantly smaller particles from the particle reservoir.  
Only after comparison with our theoretical results - when repeating the experiments was no more possible - we realized that it would have been desirable to use higher quality particles, with better characterized particle sizes. 

\begin{figure}
\begin{center}
\hspace*{-1.0cm}
\vspace*{0.3cm}
{\includegraphics[height=0.4\textwidth]{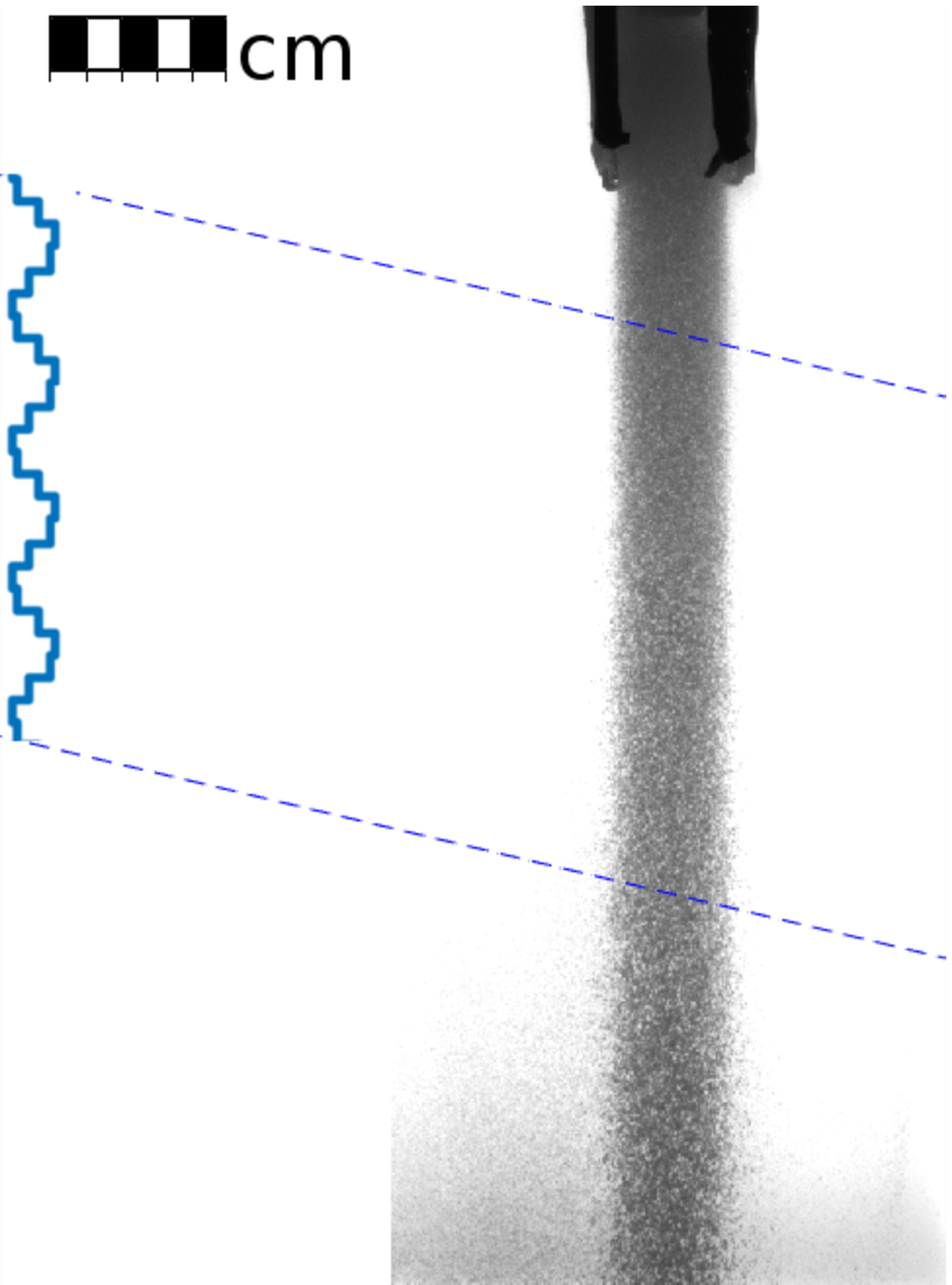}\hspace*{0.5cm}
\put(-80, 160){t=0}
\put(-100, 6){(a)}}
\put(-145,87){\rotatebox{90}{wave maker} }
{\includegraphics[height=0.4\textwidth]{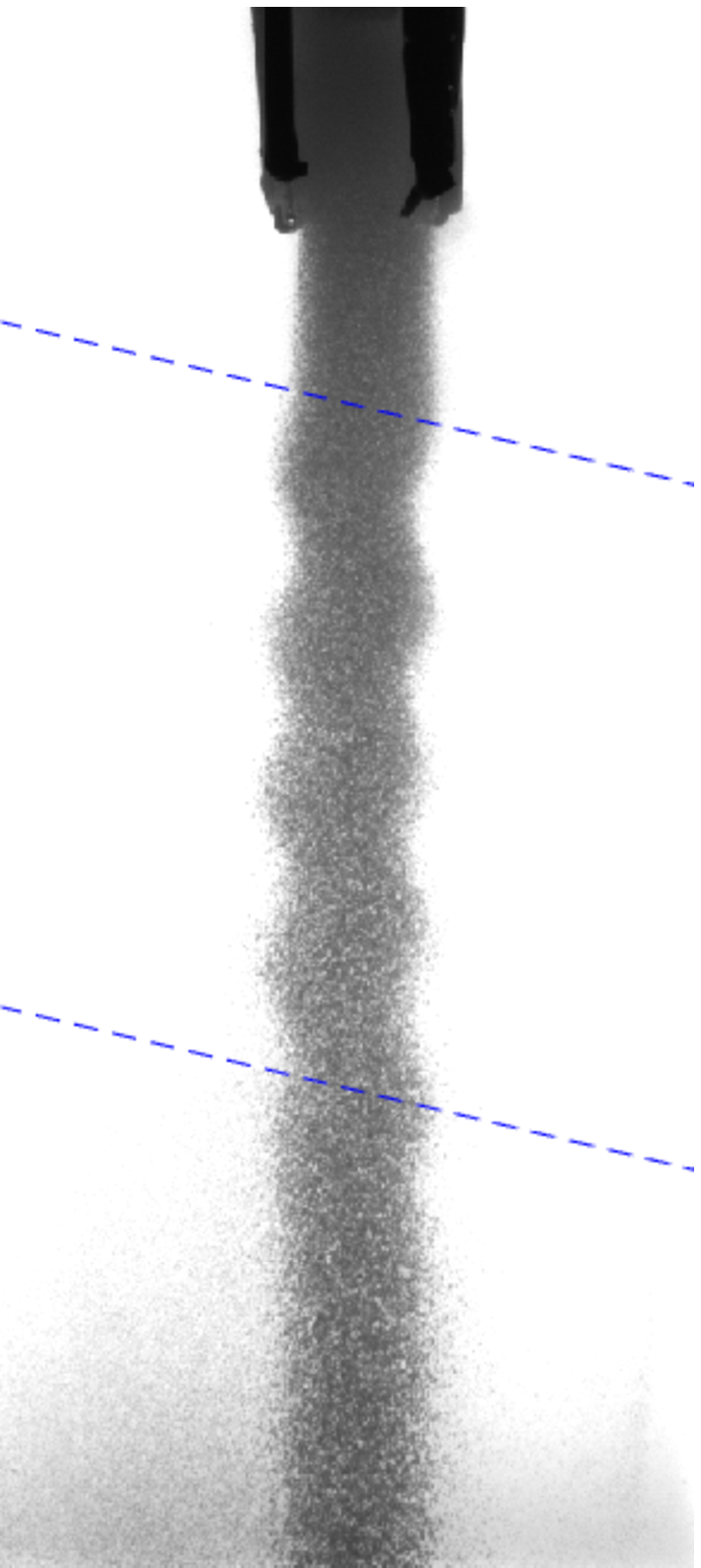}\hspace*{.5cm}
\put(-97, 160){t=6.25T}
\put(-85, 6){(b)}}
{\includegraphics[height=0.4\textwidth]{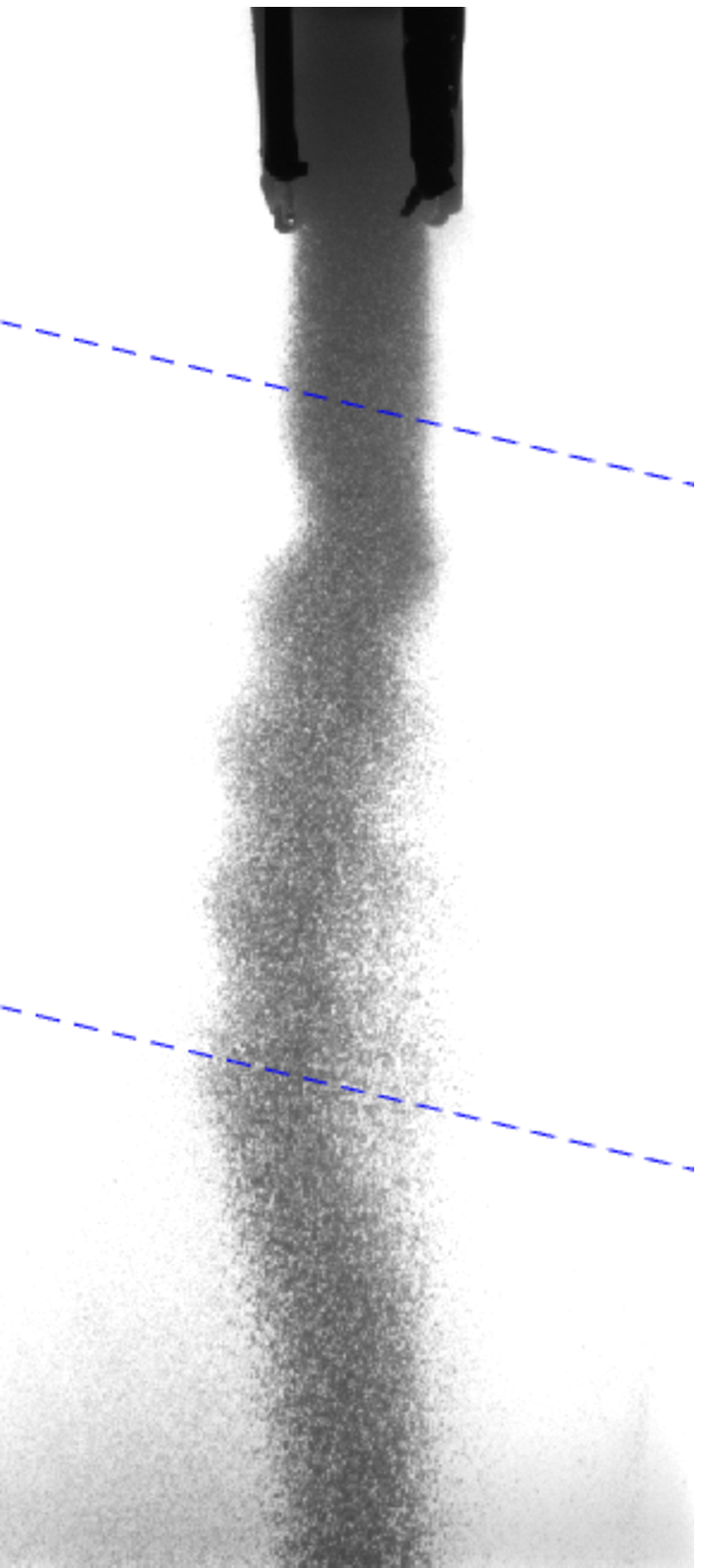}\hspace*{.5cm}
\put(-97, 160){t=12.5T}
\put(-85, 6){(c)}}
{\includegraphics[height=0.4\textwidth]{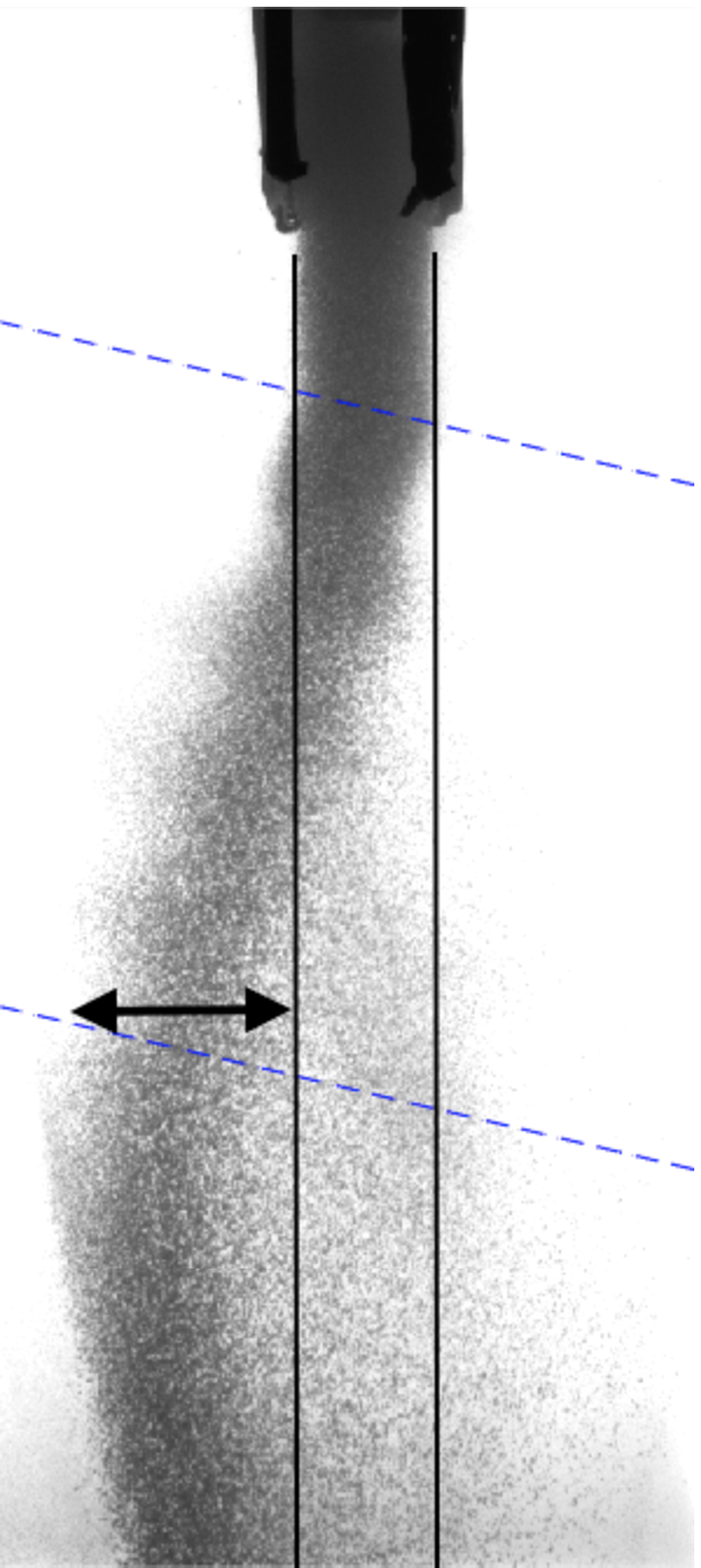}
\put(-60,63){$\delta x_{max}$}
\put(-75, 6){(d)}}
\put(-84, 160){t=37.5T}
\caption{\small{Snapshots at four instances (here, the wave period is $T=20$ s) of the sinking particle column at 18 cm from the wave maker, intersected by the wave beam between the blue dashed lines. The wave maker is switched on at time $t=0$ with fully established column of sinking particles (a). The maximum horizontal displacement, $\delta x_{max}$, of the left edge of the sinking particle column is illustrated in Fig. \ref{fig:f2}d. }}
\label{fig:f2}
\end{center}
\end{figure}

\begin{figure}
\begin{center}
\vspace{0.5cm}
\includegraphics[width=1\textwidth]{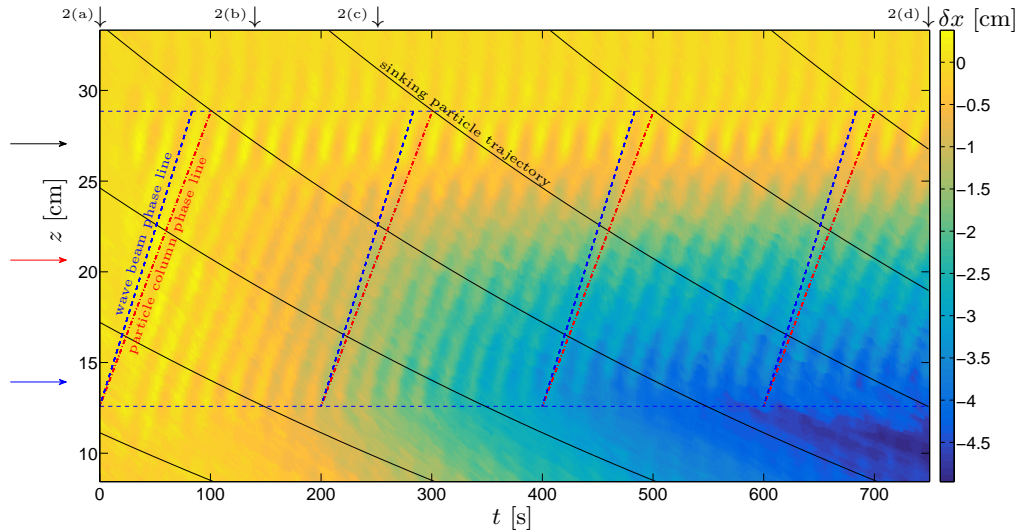}
\put(-50,195){$\delta x$ [cm]}
\put(-366 ,196){$\downarrow$}
\put(-378 ,198){\tiny 2(a)}
\put(-308 ,196){$\downarrow$}
\put(-321 ,198){\tiny 2(b)}
\put(-262 ,196){$\downarrow$}
\put(-275 ,198){\tiny 2(c)}
\put(-56,  196){$\downarrow$}
\put(-69,  198){\tiny 2(d)}
\put(-385,110){\rotatebox{90}{ $z$ [cm] }}
\put(-220,7){ $t$ [s]}
\put(-260,180){\rotatebox{-36}{\tiny sinking particle trajectory} }
\put(-359,85){\rotatebox{73}{\tiny \textcolor{blue}{ wave beam phase line} } }
\put(-354,71){\rotatebox{70}{\tiny \textcolor{red}{ particle column phase line} } }
\caption{\small{ Displacement $\delta x$ of the left edge of the sinking particle column (at 18 cm from wave maker and for $\omega_0=0.31$ rad/s), as a function of height $z$ (vertical) and time $t$ (horizontal). Particles are subject to wave beam motion between the two blue dashed horizontal lines. The four black arrows at the top indicate the corresponding snapshots in Fig. \ref{fig:f2}; the three horizontal arrows at the left indicate the vertical levels at which time series are shown in Fig. \ref{fig:f4}a. The column phase speed, $\omega_c / k_z = \omega_0/k_z + \bar{w}_p$, (indicated by four red dot-dashed lines, matching observed phase lines) differs from the wave beam phase speed, $\omega_0 / k_z = 0.195$ cm/s, (four blue dashed lines) due to a Doppler shift. The average particle sinking velociy $\bar{w}_p$ is accurately determined from the difference between $\omega_0$ and the experimentally estimated column oscillation frequency $\omega_c$, see also Fig. \ref{fig:f4}b. The curved downward-sloping lines show theoretical particle trajectories based on average sinking velocity (here $\bar{w}_{p} \approx 0.03$ cm/s at $z=20$ cm) determined from Doppler shift, with the factor 2/3 slowdown from the surface towards the bottom due to decreased density difference between stratified fluid and sinking particles (see \S \ref{particle_trajectory}).  
}}
\label{fig:f3}
\end{center}
\end{figure}


\section{Experimental results}
\label{obs}

For each experiment, the wave maker is turned on once the column of settling particles is fully established in the quiescent stratified fluid (Fig. \ref{fig:f2}a), as described in \S \ref{setup_particles}. The wave beam builds up quickly (over a few wave periods) and is in steady state throughout most of the experiment, lasting 730 seconds (in some cases 750 seconds), corresponding to $15$ to $90$ wave periods, depending on the imposed wave frequency, $\omega_0$. 
We observe that the internal wave beam perturbs the column, moving it back and forth with the wave beam motion, thus imprinting wiggles on the column snapshots in Fig. \ref{fig:f1}a and Fig. \ref{fig:f2}b-d. In addition, throughout the experiment, the particle column is slowly shifted towards the wave maker (Fig. \ref{fig:f2}b-d), which is of particular interest because the driving mechanism behind this leftward particle column motion was initially unknown. We present the displacement of the left edge of the sinking particle column, $\delta x$, for the same experiment, as a function of height $z$ and time $t$ in a contour plot in Fig. \ref{fig:f3}, and at three vertical levels in Fig. \ref{fig:f4}a. We find that the column displacement increases linearly from top to bottom of the wave beam interaction zone (Fig. \ref{fig:f2}c-d), suggesting that the settling particles are exposed to a depth-independent transport mechanism as they sink through the wave beam. Despite the overall movement to the left, we also observe that some particles are transported to the right (slightly visible in Fig. \ref{fig:f2}d). We interpret this transport to the right as a clear indication that the quasi-two-dimensional wave beam facilitates a particle transport that is non-uniform in the cross-tank direction ($y$), consistent with our theoretical explanation developed in \S \ref{theory}.
\\
\indent 
We find that the strength of the horizontal column displacement, $\delta x$, is strongly dependent on the imposed wave frequency, $\omega_0$. This is visualized in Fig. \ref{fig:f4}c, where we present the maximum displacement of the left column edge, $\delta x_{max}$, for all eleven experiments. The maximum displacement, illustrated in Fig. \ref{fig:f2}d, ranges from $0.3$ to $5$ cm to the left across the eleven experiments, with one exception for the experiment with the largest wave frequency ($\omega_0 = 0.785$), where we find a transport of $\sim 1$ cm to the right.
It is the peculiar horizontal displacement, the apparent non-uniformity in the cross-tank direction, and its strong dependence on the wave frequency, that we intend to explain by the theory developed in \S \ref{theory}. 
%


\section{Theory on mean flow generation at lateral walls}
\label{theory}
 In this theoretical part we employ small-amplitude expansions to derive exact expressions for monochromatic quasi-2D internal wave beams between two lateral walls, with viscous dissipation taken into account due to shear in the cross-beam direction and friction with the lateral walls. The exact solutions to the linearized equations, constructed in \S \ref{wave_beam}, are used in \S \ref{mean} to compute the induced mean field. In particular, we focus on the previously unrecognized mean vertical vorticity production in the stratified lateral-wall boundary layers, which drives a strong horizontal mean circulation. 
 \subsection{Internal wave beam between lateral walls}
\label{wave_beam}
For this theoretical analysis, we shall work with dimensionless variables (listed in Table \ref{t1}), employing $1/\omega_0$ as the time scale and the wave length, $L_0$, as the length scale. The velocity is non-dimensionalized by $U_0=A_0 \omega_0$, which is the along-beam velocity amplitude based on an empirical parameterisation\footnote{Based on the impermeability boundary condition of the \emph{horizontal} velocity at a vertically oriented wave maker with frequency $\omega_0$, horizontal oscillation amplitude $A_0$ and vertical wave length $L_z$, one may expect the generated along-beam velocity amplitude to be $A_0 \omega_0/\cos \theta + \mathcal{O}(A_0/L_z)$. 
However, experiments and simulations by \cite{Me10} have revealed that the ratio of observed along-beam velocity over the theoretical value $ (A_0 \omega_0 /\cos \theta)$ typically falls in between $\cos \theta$ and $\cos^2 \theta$. We use the upper bound parameterization, $U_0 = A_0 \omega_0$, for our non-dimensionalization. } for a vertically oriented, horizontally oscillating wave maker.
\begin{figure}
\begin{center}
\hspace*{-1cm}
\includegraphics[width=1.1\textwidth]{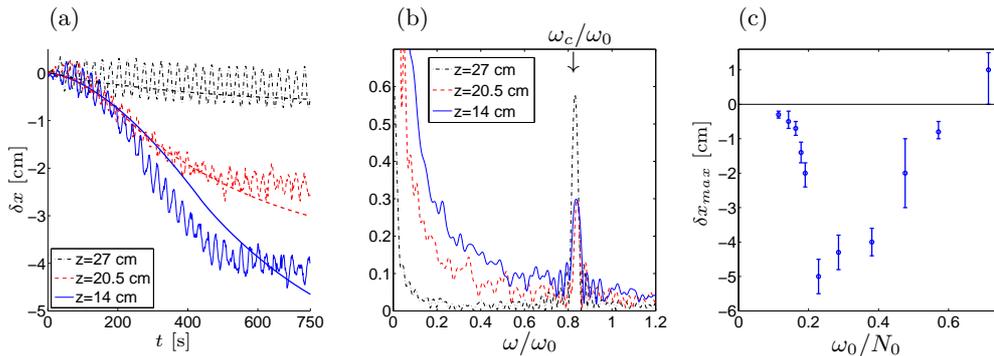}
\put(-400, 121){(a)}
\put(-270, 121){(b)}
\put(-142, 121){(c)}
\put(-230, -1){$\omega/\omega_0$}
\put(-360, 0){\scriptsize $t$ [s]}
\put(-415, 50){\scriptsize \rotatebox{90}{$\delta x$ [cm]}}
\put(-216, 115){ $\omega_c/\omega_0$}
\put(-209, 105){ $\downarrow$}
\put(-110, -2){ $\omega_0/N_0$}
\put(-162, 45){ \scriptsize \rotatebox{90}{$\delta x_{max}$ [cm]}}
\caption{\small{(a): Time series of the horizontal displacement $\delta x$ of the left edge of the sinking particle column for the experiment with $\omega_0 = 0.31$ rad/s at three heights, near the top of the wave beam (black), half way through (red) and near the lower edge of the wave beam (blue). The corresponding simulated mean column displacements (see \S \ref{theory}) are superimposed (smooth lines). (b): Corresponding spectra of experimental displacement time series, clearly peaking at Doppler shifted particle column frequency $\omega_c/\omega_0 = 0.84 \pm 0.02 $, which gives an accurate sinking velocity estimate, $\bar{w}_p = 0.33 \pm 0.04$ mm/s. (c) Maximum horizontal displacement of the sinking particle column at the end of the experiments, for 11 different wave frequencies, all with $N_0 = 1.1$ rad/s. }}
\label{fig:f4}
\end{center}
\end{figure}

Denoting the dimensionless coordinates also by $(x,y,z)$, we consider a linearly stratified Boussinesq fluid with scaled \BV frequency $N=N_0/\omega_0= 1/\sin \theta$ in an infinite domain between two lateral walls at $y=\pm l_y=\frac{W}{2L_0}$ and for $x\geq 0$.\\
The equations governing the dimensionless velocity field ${\bf u}=(u,v,w)$, buoyancy $b$, and pressure $p$ of the Boussinesq fluid are given by
\begin{equation}
\begin{split}
{\bf u}_t+\epsilon \ \left( {\bf u} \cdot  \nabla \right)  {\bf u}=\,& - \nabla p+\delta^2 \Delta {\bf u} + {\bf \hat{z}}b, \qquad b_t +\epsilon \ {\bf u}  \cdot  \nabla b=  -N^2 w, \qquad \nabla  \cdot {\bf u}= 0.
\label{ge1}
\end{split}
\end{equation}
Here, $\epsilon=\frac{U_0}{\omega_0 L_0}\ll 1$ is the Stokes Number, and $\delta=\frac{d_0}{L_0}\ll 1$ is the thickness of the Stokes boundary layer, $d_0=\sqrt{\nu / \omega_0}$, scaled by wave length $L_0$, and where $\nu=1$ mm$^2$/s is the kinematic viscosity. We solve (\ref{ge1}) with no-slip boundary conditions, ${\bf u}={\bf 0}$ at $y=\pm l_y$ by expanding the velocity vector ${\bf u}$ in the small parameters $\delta$ and $\epsilon$, 
\begin{equation}
 {\bf u}(t)={\bf u}_0 e^{-i t}+\delta{\bf u}_{1} e^{-i t}+ {\bf \bar{u}}(\epsilon t)+\mathcal{O}(\delta^2,\epsilon^2) ,
\label{u_expansion}
\end{equation}
 and similarly for buoyancy $b$ and pressure $p$, valid for the time range $t \in [0, \mathcal{O}(\epsilon^{-1})]$. Here, and in the following, physical quantities are always the real part of the presented expression. The overbar denotes the components which vary only over the slow time, $\tau=\epsilon t$, i.e. $\partial_t {\bf \bar{u}}(\tau)\in\mathcal{O}(\epsilon)$. These slowly-varying velocity components,  with $\bar{{\bf u}}={\bf 0}$ at $\tau=0$, are  induced through Reynolds stresses of the time-periodic velocity field ${\bf u}_0(t)+\delta{\bf u}_{1}(t) $. It turns out that it is strictly necessary to include the $\mathcal{O}(\delta)$ transversal velocity, $v_1$, as it contributes to the leading order Reynolds stress near the lateral walls. We also assume that $\delta l_y^{-1}\ll 1$, that is to say, the dimensional half width of the domain, $\frac{W}{2}= l_y L_0$, is much larger than the Stokes boundary layer width, $d_0$. 
 
  \begin{center}
 \begin{table}
\begin{tabular}{lll}
{\bf  Parameter       }                           &  {\bf Dimensional, in experiments }	                                    & {\bf Non-dimensionalized } \\
 Wave frequency 					 		& $\omega_0=0.125-0.785$ rad/s			                                   	& $1$ 				\\
Along-beam velocity amplitude      & $U_0= A_0 \omega_0 = 1 - 7 $ mm/s	 & 1      \\
Cross-beam wave length			            		    & $L_0=\cos \theta L_z=2.76-3.88$ cm									& 1						\\
Vertical wave length					    & $L_z=3.9$ cm									            		& $1/\cos\theta = 1.006 - 1.435$			\\
Angle $\theta=\sin^{-1}[N_0/\omega_0]$	    & $\theta=0.11 - 0.79$   rad							        		& $\theta$, same values		\\
\BV frequency			 					& $N_0=1.1$ rad/s 			                                    	& $N=N_0/\omega_0 = 1/\sin \theta =1.4 - 8.8$\\
Amplitude of wave maker forcing				& $A_0 = 9$ mm											        	& $U_0 \cos \theta / (L_0 \omega_0) = \epsilon \cos \theta = 0.23 $					\\
Width of the tank (in $y$-direction)		& $W=2 l_y L_0=17.0$ cm							        			& $2l_y=4.4 - 6.2	$			\\
Height of wave maker and beam		& $4.17 L_z = 16.25$ cm							        			& $2h=4.17/\cos\theta =4.2 - 5.9	$			\\
Boundary layer thickness 					&  $d_0=\sqrt{\nu/\omega_0}=1.1 - 2.8$ mm   				    			&  $\delta=d_0/L_0=0.04 - 0.08 \ll 1$		\\
Decay rate due to internal shear & $\beta_1 /L_0 = 0.02 - 0.08 $   cm$^{-1}$							        	& $\beta_1 = 0.07 - 0.2 $							\\
Decay rate due to wall friction & $\beta_2 /L_0 = 0.005 - 0.02 $   cm$^{-1}$							        	& $\beta_2 = 0.02 - 0.06 $							\\
Stokes Number	 at wave maker				& $\epsilon=U_0 / (\omega_0 L_0) = 0.23 - 0.33$ 			                                & $\epsilon$, same value	 \\
Stokes Number	 at particle column				& $\epsilon_{18cm} = 0.05 - 0.15 \ll1$ 			                                & same values	 \\
~
\end{tabular}
\caption{ 
Parameter values of the experiments and their corresponding non-dimensional values. Note that the Stokes number $\epsilon_x = \epsilon \exp[-(\beta_1+\beta_2) x/L_0 ]$, quantifying the weak non-linearity of the wave field, decreases strongly with distance to the wave maker. Both $\epsilon$ and $\delta$ are sufficiently small for our perturbation analysis to be valid.  \\ }
\label{t1}
\end{table}
\end{center}

\subsection*{2D wave beam}
\begin{figure}
    \centering
 \hspace*{-1.0cm}
    \includegraphics[width=\textwidth]{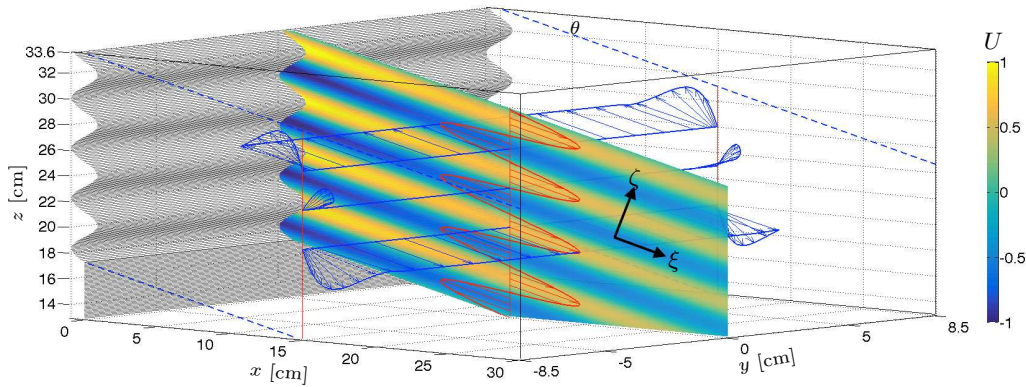}
    \put(-22,135){$U$}
    \put(-180,141){ \scriptsize $\theta$}
     \put(-300,12){ \scriptsize \rotatebox{-5}{$x$ [cm]}}
    \put(-390,70){ \scriptsize \rotatebox{90}{$z$ [cm]}}
    \put(-120,13){  \scriptsize \rotatebox{10}{ $y$ [cm]}}
   \put(-140,53){$\xi$}
   \put(-155,84){$\zeta$}
    \caption{Snapshot of the theoretical along-beam velocity,  $U$, in the center plane, $y=0$, with parameter values corresponding to the experiment with $\omega_0 = 0.31$ rad/s. The velocity vector ${\bf u}_0 = [u_0, 0, w_0]$ is presented at $x=15$ cm along three cross-tank transects ($z=19.9,\ 22.95,\ 26$ cm, blue arrows) and along a vertical transect in the center plane (red arrows). For visualization purposes, the magnitude of the vector ${\bf u}_0$ is elongated by a factor $3$ in the along-phase propagation direction, $\zeta$. Note that near the lateral walls, the velocity vector ${\bf u}_0$ is \emph{not} aligned with the along-beam direction, $\xi$. As a result, the in-product of ${\bf u}_0$ with $\nabla u_0$ (pointing along the phase propagation direction, $\zeta$) is  non-negligible (see also Fig. \ref{fig:f6}a), causing strong mean flow generation near the lateral walls (shown in Fig. \ref{fig:f7}).  }
    \label{fig:f5}
\end{figure}

Motivated by the experiments, we consider the internal wave energy propagation to be downwards along coordinate $\xi=x \cos \theta - z \sin \theta $, implying that the phase propagation is upwards along $\zeta=x \sin \theta +z \cos \theta $. The leading order velocity field of transversally uniform wave beams solving (\ref{ge1}) at $\mathcal{O}(\delta^2, \epsilon^0)$-accuracy with free-slip condition at the lateral walls can be expressed as
 \begin{equation}
 [u_0,v_0,w_0]=[\cos \theta, 0, -\sin \theta] U , \quad \mbox{ with } \quad U = \frac{1}{2\pi} \int_0^{\infty}  \hat{U}(k) e^{i k \zeta -\left(\beta_1 + \beta_2\right) \frac{x}{\cos \theta } -i t} dk.
 \label{uvw0}
 \end{equation}
 Here, $\beta_1 $ and $\beta_2$ are the viscous decay rates corresponding to internal shear dissipation and friction with the lateral walls, respectively. The spectrum $\hat{U}(k)$ of the along-beam velocity component is presumed to vanish for negative wave numbers, $k<0$, because the phase propagation is primarily along positive $\zeta$\footnote{It is known \citep{Me10, Be18b} that the upper and lower edges of the wave maker act as line sources, also radiating waves upwards. These upward-propagating waves, hardly visible in the experimental velocity field (Fig. \ref{fig:f1}a), are not relevant for the dynamics discussed here.}. 
 Replicating the laboratory experiment, we consider the along-beam velocity spectrum 
 \begin{equation}
 \hat{U}(k)= \frac{\sin\left[ h \cos \theta( k - 2\pi) \right] }{\pi ( k-2\pi)}, 
 \label{Uk}
\end{equation}
 generated by a wave maker of height $2h$ with vertical wave number $2\pi \cos \theta$ at $x=0$, as depicted in Fig. \ref{fig:f5}.
 \\
 \indent 
 The along-beam decay rate due to shear in the cross-beam direction $\zeta$ can be derived in various ways (see \S6 in \cite{Vo03} for an overview) and is given by $\beta_1 = \delta^2 \tan \theta k^3/2$. For the laboratory experiments with the cross-beam widths $2h \cos \theta = 4.17$, the spectrum $\hat{U}(k)$ peaks sharply at the imposed wave number $2\pi$, justifying the use of the simplified internal shear decay rate $\beta_1 = 4 \pi^3 \delta^2 \tan \theta $ for $k=2\pi$. The decay rate $\beta_2$ obviously vanishes for free-slip conditions at the lateral walls, and is determined next upon imposing no-slip boundary conditions.

\subsection*{Lateral boundary layer}

Following the analysis by \cite{BM16,Be18}, we consider the momentum equations for $u_0$ and $w_0$ in stretched transverse coordinate  $\eta=\delta^{-1} y$:
 \begin{equation} 
-i u_{0}= -p_{0_x}+u_{0_{\eta\eta}}, \quad          i \cot^2 \theta \ w_{0}=-p_{0_z}+w_{0_{\eta\eta}}.
\label{2m}
\end{equation}
Imposing no-slip boundary conditions at the walls, $\eta=\pm \delta^{-1}l_y$, and interior velocity field (\ref{uvw0}), gives
 \begin{equation}
\begin{split}
\,&  u_0= \cos \theta \left(1- E^x \right) U, \  \quad \qquad \qquad  E^x(\eta)=\frac{\cosh[ i^{-\frac{1}{2}} \eta ]}{\cosh[ i^{-\frac{1}{2}} \delta^{-1}l_y] },  \\
\,& w_0=- \sin \theta\left(1- E^z \right) U, \ \qquad \qquad E^z(\eta)=\frac{\cosh[   i^{\frac{1}{2}} \cot \theta\  \eta]}{\cosh[  i^{\frac{1}{2}} \cot \theta \ \delta^{-1}l_y]}.
\label{uw}
\end{split}
\end{equation}
The non-planar orbital structure of this along-beam velocity vector, $[u_0, w_0]$, is illustrated in Fig. \ref{fig:f5}. The velocity vector $[u_0, 0, w_0]$ is divergent near the lateral walls due to the presence of stratification. Hence, by the continuity equation $ u_{0_x}+w_{0_z}=- v_{1_\eta}$ at $\mathcal{O}(\delta^0)$ in stretched coordinate ($\eta$), and using the impermeability boundary conditions $v_1=0$ at $y = \pm l_y$, we find the $\mathcal{O}(\delta)$ transversal velocity
\begin{equation}
v_1=\sin \theta  \left(  i^{\frac{1}{2}} \cos \theta \frac{\sinh \left[  i^{-\frac{1}{2}}  \delta^{-1} y \right]}{\cosh \left[  i^{-\frac{1}{2}}  \delta^{-1}l_y \right] }   -    i^{-\frac{1}{2}} \sin \theta \frac{\sinh \left[ i^{\frac{1}{2}} \cot \theta \  \delta^{-1} y \right]}{\cosh \left[ i^{\frac{1}{2}} \cot \theta \ \delta^{-1}l_y \right] }  - i^{\frac{1}{2}} e^{i \theta }  \frac{y}{l_y} \right)  U_{\zeta}.
\label{v1}
\end{equation}
The slow $\mathcal{O}(l_y^{-1})$-decay of $v_1$ towards the interior, $  - i^{\frac{1}{2}} \sin \theta e^{ i \theta } U_{\zeta} y/l_y $, requires an additional viscous attenuation factor, $\exp[-\beta_2 \xi]$ with decay rate $\beta_2 = \Re[i^{-\frac{1}{2}} \sin \theta e^{i \theta} \delta l_y^{-1} k]$, to satisfy the $\mathcal{O}(\delta)$-continuity equation \citep{BM16}. Again, we may replace $k$ by the dominant wave number, $2\pi$. Despite the relatively thin boundary layers, we find that the friction with the lateral walls almost doubles the decay at $60$ cm from the wave maker for the experiment depicted in Fig. \ref{fig:f1}.

\subsection{Induced Eulerian mean flow}
\label{mean}
In this section, we construct the so-called induced mean flow, ${\bf \overline{u}}=[\bar{u},\bar{v},\bar{w}]$, generated through the time-averaged Reynolds stresses at $\mathcal{O}(\epsilon)$. 
Balancing time-independent terms in the buoyancy equation at $\mathcal{O}(\epsilon)$, one readily finds the vertical induced mean flow
 \begin{equation}
 \bar{w}=
 -\frac{\epsilon }{N^2} <\Re[{\bf u}^0]\cdot \Re[\nabla b_0]> 
 \ =\ -\frac{\epsilon }{2N^2} \Re[{\bf u}^0\cdot \nabla b_0^*]
 \ =\  \frac{\epsilon}{2} \Im[{\bf u}^0  \cdot \nabla w_0^*],
 \label{barw}
 \end{equation}
 where ${\bf u}^0=[u_0,\delta v_1, w_0]$, $b_0=-i N^2 w_0$, and $<\cdot >$ stands for time-averaging over one wave period, $2\pi$. This weak (order-$\epsilon$) induced mean vertical flow (in Eulerian framework) is exactly balanced by the vertical Stokes Drift of the wave beam velocity field, ${\bf u}^0$, which we show in the appendix, \S \ref{stokes_drift}. Consequently, net mass transport can only take place in the horizontal plane, and we thus focus on the horizontal induced circulation. 
\begin{figure}
    \centering
    \begin{minipage}{.7\textwidth}
    \hspace*{-1.8cm}
    \includegraphics[height=0.37\textwidth]{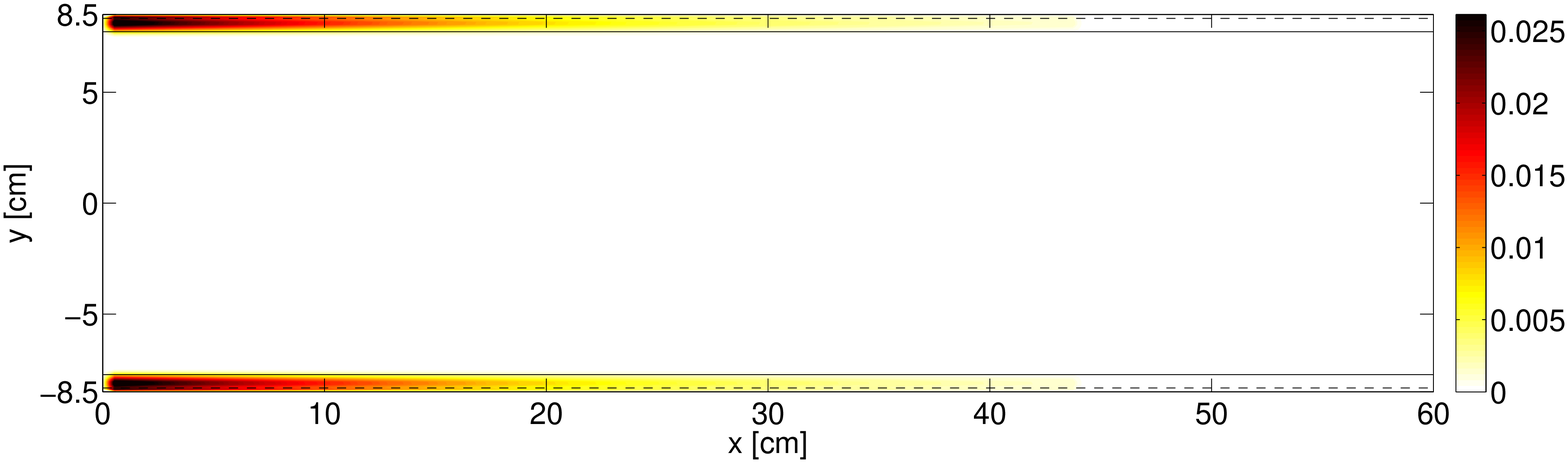}
    \put(-48,98){\scriptsize $\bar{F}$ [m/s$^2$]}
    \put(-350,90){(a)}
    \end{minipage}
    \begin{minipage}{0.29\textwidth}
    \hspace*{0.7cm}
    \includegraphics[width=0.95\textwidth]{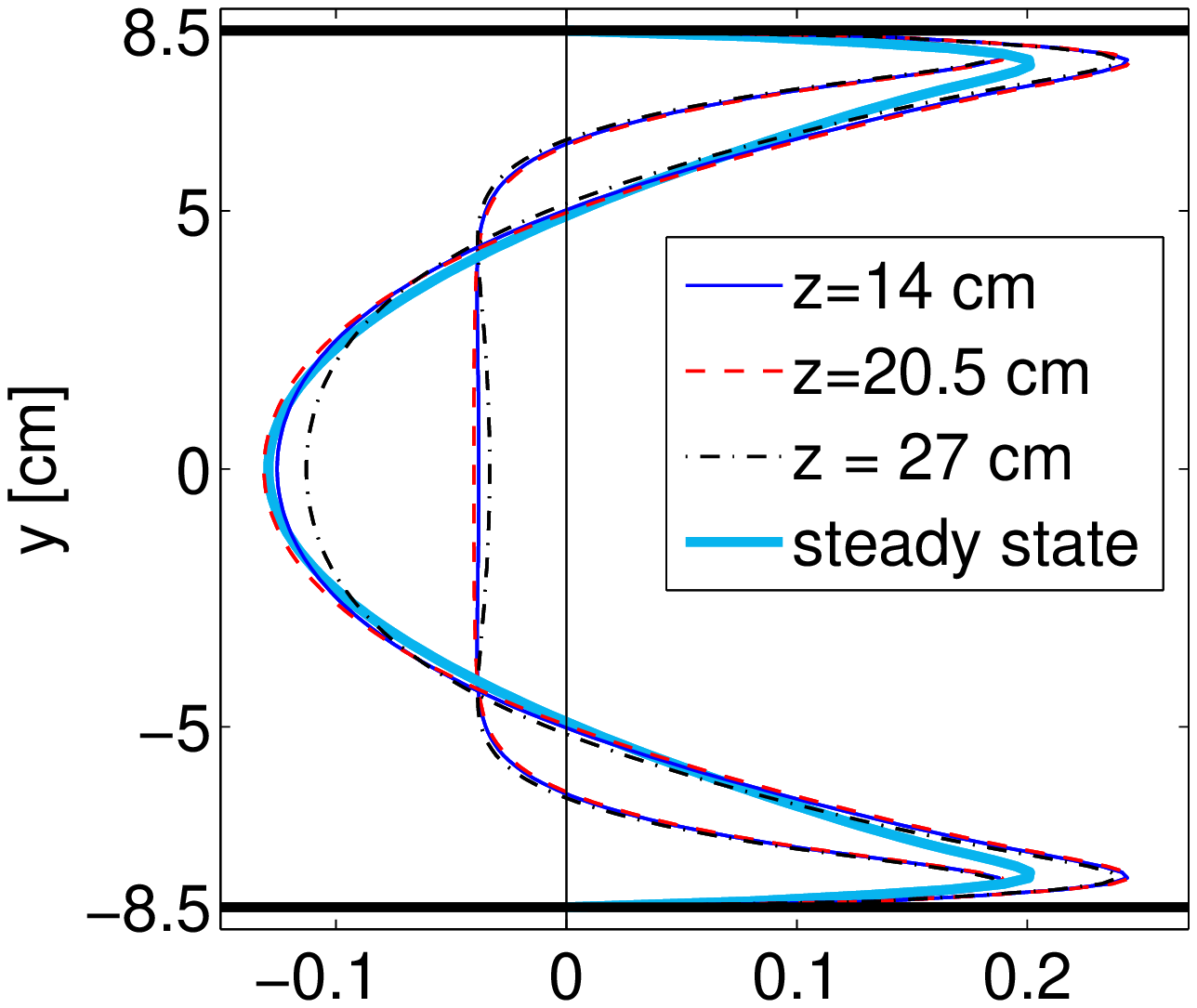}
    \put(-77, -3){ $U_0 \bar{u}$ {\scriptsize [mm/s]}}
    \put(-118, 90){(b)}
    \put(-71,35){\tiny $t=5$T}
    \put(-56,27){ \tiny $t=37.5$T}
    \end{minipage}
    \caption{(a) Spatial structure of along-wall Reynolds stress, $\bar{F}$, for parameter values corresponding to the experiment with $\omega_0=0.31$ rad/s, at $z=20.5$ cm, vanishing almost everywhere except within a distance $(1+\cot \theta) d_0 =0.78$ cm of the lateral walls (indicated by solid lines). This distance is the sum of the homogeneous boundary layer thickness, $d_0 = \sqrt{\omega_0/\nu}=0.18$ cm (dashed lines) and the (stratified) boundary layer thickness of the vertical velocity component, $d_0 \cot\theta = 0.69$ cm. (b) Simulated mean flow in $x$-direction,  $U_0\bar{u}$, along a cross-tank transect ($y$) at three vertical levels and $x=17$ cm, at times $t=5$ T and $t=37.5$ T. The thick light blue line corresponds to the theoretical steady state mean velocity, derived in \S \ref{steady_state_induced_mean}.}
    \label{fig:f6}
\end{figure}
\\
\indent 
 Using a Helmholtz decomposition, the horizontal induced mean velocity field can be split into $[\bar{u},\bar{v}] = [ \bar{\Psi}_y + \bar{\phi}_x, -\bar{\Psi}_x + \bar{\phi}_y]$, where $\bar{\Psi}$ and $\bar{\phi}$ are the stream function and flow potential, respectively. The flow potential is set by the vertical mean velocity, $\bar{w}$, through the continuity equation, $\bar{\phi}_{xx}+ \bar{\phi}_{yy} = - \bar{w}_z$, and is weak at all times. Whereas the flow potential, $\bar{\phi}$, investigated by \cite{KC01, TA03} among others, may be relevant for truly two-dimensional wave fields (where $\bar{\phi}_y=0$), it is typically secondary to the  vortical flow described by the stream function, $\bar{\Psi}$, in the quasi-two-dimensional and three-dimensional configurations \citep{Be18b}. The mean vortical flow associated with $\bar{\Psi}$ can persistently accumulate energy until a strong large scale circulation is established. The slow evolution equations of the potentially strong horizontal vortical induced mean flow associated with stream function $\bar{\Psi}$, and no-slip boundary conditions at the tank boundaries, $x=0,l_x$ and $y=\pm l_y$, are given by 
 \begin{equation}
 \begin{split}
 & \Delta_h \bar{\Psi}_{\tau} =  \epsilon^{-1}\delta^2 \Delta_h^2 \bar{\Psi} +  \bar{F}_y\qquad \mbox{ with } \qquad
 \bar{F}=-<\Re[ {\bf u}^0 ]\cdot \Re[ \nabla u_0]>, \\
 & \bar{\Psi}(x,\pm l_y)=\bar{\Psi}_y(x,\pm l_y)=0, \qquad   \bar{\Psi}(0,y)=\bar{\Psi}_x(0,y)=\bar{\Psi}(l_x,y)=\bar{\Psi}_x(l_x,y)=0,
 \label{Psi}
 \end{split}
 \end{equation}
 Here, $\Delta_h$ is the horizontal Laplace operator, and $\tau = \epsilon t$ is the slow time over which the time-averaged $\mathcal{O}(\epsilon)$-Reynolds stress, $\epsilon \bar{F}$, acts as a $\mathcal{O}(1)$-source of mean vertical vorticity. We also decoupled the three-dimensional induced mean flow problem into independent two-dimensional planar problems by neglecting shear in the vertical direction. This is an appropriate simplification, because lateral wall friction turns out to dominate induced mean flow damping. \\
A lengthy but straightforward small-amplitude expansion simplifies the mean Reynolds stress, $\bar{F}$, to
 \begin{gather}
 \bar{F} =  -\frac{1}{2} \Re \left[ {\bf u}^0 \cdot \nabla u_0^* \right] = 
 -\frac{ \sin \theta \cos^2 \theta }{2} \sum \limits_{n=1}^5 \Im \left[ a_n 
 \frac{\cosh\left[ c_n  \delta^{-1} y \right] }
 {\cosh \left[ c_n \delta^{-1} l_y \right] } \right] 
 \Im \left[ U U_{\zeta}^* \right]+\mathcal{O}(\delta l_y^{-1}) \label{Rx} \\
 \mbox{ with } \ 
 a_1 = -1  - \frac{ i e^{-i \theta}}{\cos \theta}, \quad 	
 a_2 =  1,   \quad   
 a_3 =  \tan \theta, \quad  
 a_4 = - 1	,   \quad      
 a_5 =  1+i,\quad   \nonumber \\
  \ \qquad 
  c_1 = i^{-1/2},		\quad			 
  c_2 = i^{1/2} \cot \theta,  \quad   
  c_3 = i^{-1/2} \left( \cot \theta+1 \right), \quad
  c_4 = i^{ 1/2} \left( \cot \theta+1 \right),  \quad 
  c_5 = \sqrt{2}. \nonumber
 \end{gather}
Here, we have absorbed $\Re[ U U_{\zeta}^*]$ - slightly non-zero due to along-beam viscous dissipation - in $\mathcal{O}(\delta l_y^{-1})$. The spatial structure of the mean Reynolds stress, $\bar{F}$, which is practically zero everywhere except in a neighborhood of $d_0(1+\cot \theta)$ near the lateral walls, is presented in Fig. \ref{fig:f6}a.  \\
\indent 
Scaling analysis also reveals that the forcing due to transversal advection, $<\Re[{\bf u}^0]\cdot \Re[\nabla v_1]>$, is at most $\mathcal{O}(\epsilon \delta)$, hence it does not contribute to the leading order Reynolds stresses at $\mathcal{O}(\epsilon)$. Note also that in the interior, the largest Reynolds stresses are $\mathcal{O}(\epsilon \delta l_y^{-1})$ due to viscous dissipation by the lateral walls, and $\mathcal{O}(\epsilon \delta^2)$ due to internal shear dissipation. For the laboratory experiments, we find that Reynolds stresses in the interior contribute at most 3.5\% of the original wave maker forcing ($u_0 u_{0_{\xi}} \sim (\beta_1+\beta_2)\epsilon$ relative to $u_{0_t} \sim \mathcal{O}(1)$), whereas near the lateral walls, the Reynolds stresses (\ref{Rx}) reach up to $25$ \%. We solve Eq. \ref{Psi} numerically on a rectangular grid, as explained in the appendix, \S \ref{numerical_sim}. Snapshots of the simulated induced mean flow in three horizontal planes are shown in Fig. \ref{fig:f7}, with three corresponding cross-tank transects in Fig. \ref{fig:f6}b. Simulated induced mean velocities and associated particle column displacements for different wave frequencies are presented in Fig. \ref{fig:f8}.

 \begin{figure}
\centering
\begin{minipage}{.99\textwidth}
 \hspace*{-0.5cm}
 \vspace{0.cm}
  \includegraphics[width=1.0\textwidth]{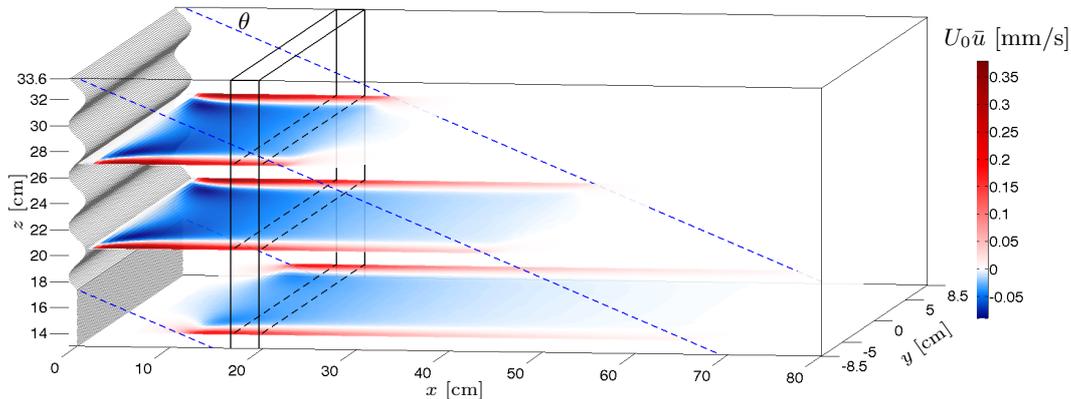} 
    \put(-33,135){$U_0\bar{u}$ [mm/s]}
    \put(-297,141){$\theta$}
     \put(-230,3){ \scriptsize $x$ [cm]}
    \put(-386,65){ \scriptsize \rotatebox{90}{$z$ [cm]}}
    \put(-56,13){  \scriptsize \rotatebox{35}{ $y$ [cm]}}
    \end{minipage}
\caption{Visualisation of the along-wall induced velocity, $U_0\bar{u}$, (in mm/s) at time $5T=100$ s, in three horizontal planes ($z=14$, $20.5$ and $27$ cm). The rectangular region where particles sink, centered at $x=180$ mm, is sketch by the black box, and the wave beam motion is confined to the region between the blue dashed lines.  Note that velocities in the boundary layer are a magnitude larger compared to the interior flow, and that the bottom of the plot ($z=13$ cm) does not correspond to the bottom of the tank ($z=0$).}
\label{fig:f7}
\end{figure}

\subsection{Steady state induced mean flow}
\label{steady_state_induced_mean}
The objective is to derive a simple expression for the horizontal induced mean velocity. We simplify the problem by assuming the domain to be infinitely long in the along-tank $x$-direction, with spatial variance of the Reynolds stress $\bar{F}$ only in the cross-tank $y$-direction. By Eq. (\ref{Psi}), the Reynolds stress, $\bar{F}_y$, is balanced by $\delta^2 \Delta^2 \bar{\Psi}$, reducing to
$$\delta^2 \bar{\Psi}_{yyyy}= -\epsilon  \bar{F}_y \qquad \Leftrightarrow \qquad \delta^2 \bar{u}_{yyy} = -\epsilon \bar{F}_y.$$
Direct integration, and incorporating the boundary condition $\bar{u}=0$ at $y=\pm l_y$, symmetry around $y=0$ and zero net mass transport, $\int \limits_{-l_y}^{l_y} \bar{u} dy=0$,  gives
\begin{equation}
\bar{u} =  \epsilon \frac{ \sin \theta \cos^2 \theta}{4 }\Im \left[ U U_{\zeta}^* \right] \left(
   \sum \limits_{n=1}^5 \Im \left[ \frac{a_n}{c_n^2} \left( 1-3\frac{y^2}{l_y^2} +
 2\frac{\cosh\left[ c_n  \delta^{-1} y \right] } 
 {\cosh \left[ c_n \delta^{-1} l_y \right] } \right) \right] 
  \right).
  \label{steady_state_mean}
   \end{equation}

We present this steady state induced mean velocity in Fig. \ref{fig:f6}b along a transect from wall to wall at $x=18$ cm, and in Fig. \ref{fig:f8}b at the center of the particle column, $(x,y)=(18,0)$ cm, as a function of wave frequency $\omega_0$. Note that some simulated mean velocities exceed the theoretical steady state mean flow. This reveals that our assumption on the forcing $\bar{F}$ being uniform along $x$-direction slightly underestimates the simulated steady state mean velocity. \\
\noindent Dimensionalising expression \ref{steady_state_mean}, and approximating the sum by 
$(\sin \theta+1)/(2 \cos^2 \theta)$, we find a characteristic value for the interior induced mean return flow:
\begin{equation}
U_0\bar{u}(y=0) \approx -\frac{\pi}{4} \sin \theta(1 + \sin \theta) \frac{U_0^2}{L_0 \omega_0 }.   
\label{baru_dim}
\end{equation}

Interestingly, the magnitude of the steady state induced mean flow is independent of viscosity ($\nu$) and of the distance between the lateral walls $(W)$. This implies that the steady state mean flow does \emph{not} vanish in the limit of zero viscosity whilst the driving Reynolds stress does vanish for zero viscosity. It is the time scale over which the steady state is reached that approaches infinity as viscosity vanishes or when the width of the domain goes to infinity. 
\\ 
In \S \ref{con} we consider oceanic conditions in which the interior return flow, Eq. \ref{baru_dim}, may be relevant.


\section{Comparison between theoretical and experimental results} 
\label{comparison}
This section is devoted to a detailed comparison between the experimental results in \S \ref{obs}, and the theory derived in \S\ref{theory}, providing strong evidence that the observed particle displacement is facilitated by the previously unrecognized internal wave lateral-wall streaming. 
\\
\indent
The theoretical results derived in detail in \S \ref{theory} can be summarized as follows: 

{\addtolength{\leftskip}{10 mm}
\addtolength{\parindent}{-10 mm}
 (i) The lateral walls modify the otherwise two-dimensional internal wave field. The most noticeable effect of the lateral walls on the linear internal wave beam dynamics is the additional viscous attenuation due to boundary friction, as discussed in \S \ref{obs} and shown in Fig. \ref{fig:f1}b. 

}
{\addtolength{\leftskip}{10 mm}
\addtolength{\parindent}{-10 mm}
 (ii) Whereas in the interior the linear wave velocity vector is practically orthogonal to the phase direction, resulting in negligible non-linear terms, this is not the case in the lateral boundary layers (Fig. \ref{fig:f6}a). The stratification causes differences in the boundary layer \emph{thickness} for vertical and horizontal velocity components, producing strong non-linearities near the lateral walls. The non-linear Reynolds stresses, driving a mean flow, are strongest for strongly inclined beams, when vertical and horizontal velocity components are of similar magnitude. The strong dependency on the wave frequency $\omega_0$ is a manifestation of the underlying dependency on the wave beam slope, $\tan \theta = \omega_0/\sqrt{ N_0^2 - \omega_0^2}$. 

}
{\addtolength{\leftskip}{10 mm}
\addtolength{\parindent}{-10 mm}
 (iii) The directly forced mean flow in the lateral boundary layers and in the direction of the horizontal beam-propagation is balanced by a return flow through the interior. The return flow is initially nearly uniform in the interior if the stretch of along-wall forcing exceeds the wall-to-wall distance. 

}

\begin{figure}
\begin{center}
\hspace*{-1.0cm}
\includegraphics[width=1.1\textwidth]{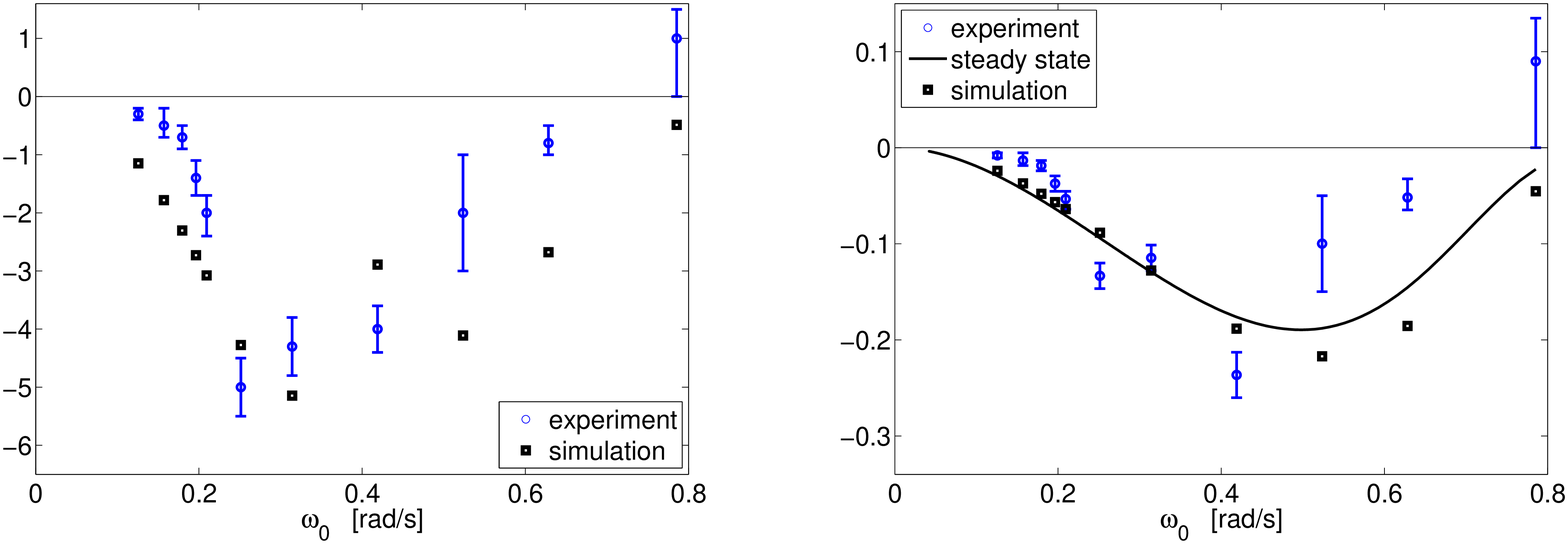} 
\put(-220,50){\rotatebox{90}{ $\bar{u}$ [mm/s]}}
\put(-420,50){\rotatebox{90}{ $\delta x_{max}$ [cm]}}
\put(-420,120){(a)}
\put(-220,120){(b)}
\caption{\small{(a): Blue dots with error margin show maximum observed displacement of sinking particle column at lower edge of interaction zone with wave beam. Black dots: simulated column displacement, where we used the Doppler shifted sinking velocity to determine how long particles take to sink through the wave beam with height $16.25$ cm. (b): In solid line, the theoretical steady state induced mean flow (see equation~\ref{baru_dim}) is compared with simulated induced mean flow at $x=165$ cm (black squares). The blue dots show the observed column displacement divided by its residence time in the wave beam interaction zone, with a maximum of $365$ seconds.   }}
\label{fig:f8}
\end{center}
\end{figure}

    If the experimentally observed particle transport is facilitated by streaming at the lateral walls, then the observations should be consistent with the described theoretical results. We can compare temporal evolution, spatial patterns and $\omega_0$-dependence of the particle displacement. In Fig. \ref{fig:f4}a we compare simulated and experimental evolution of the column edge at three vertical levels of the interaction zone of sinking particle column and wave beam. Whereas we find very good agreement at the top level, we attribute discrepancies at the middle and lower levels primarily to uncertainties in the residence time of particles within the interaction zone. Fig. \ref{fig:f7} illustrates that the directly forced flow at the lateral walls (in red) and the interior return flow (in blue) strongly decay with distance to the wave maker. Particles subject to the return flow are thus accelerated as they approach the wave maker, a feedback process that we did not take into account in the numerical simulations. 
    \\
    \indent 
    While the strength of the directly forced mean flow at the lateral walls varies strongly in the cross-tank direction, the return flow is initially roughly uniform in cross-tank direction (see profile at $t=5T$ in Fig. \ref{fig:f6}b). This is consistent with the observations: the sinking particle column is shifted to the left in the interior, maintaining a sharp left edge, while the particle advection to the right near the tank's boundaries blurs the right edge of the particle column (visible in the snapshots in Fig. \ref{fig:f2}c,d). 
    \\
    \indent 
    The strongest evidence for the lateral-wall streaming mechanism generating the observed horizontal particle displacement can be seen in the matching frequency-dependency, presented in Fig. \ref{fig:f8}. Albeit large discrepancies at particular frequencies, it is the overall strong variation with $\omega_0$ that matches surprisingly well. An exception is the experiment with the largest wave frequency ($\omega_0 =0.785$ rad/s), which clearly shows transport towards the opposite direction, at odds with the boundary layer streaming hypothesis. It appears that the horizontal mean velocity field has changed sign, generating transport to the right, blurring the left column edge (not shown). It may be useful to remark that for this particular experiment, the beam slope, $\tan \theta = 0.87$, exceeded the maximum wave maker inclination, $ L_z/(2\pi a_0) = 0.69$, implying (theoretical) fluid motion \emph{through} the wave maker plates, with possibly unknown non-linear dynamics occurring at the wave maker.  \\
    \indent 
    Last but not least, we also find reasonable agreement between theoretical induced mean velocities and velocities inferred from the observed particle displacement (Fig. \ref{fig:f8}b).

\section{Discussion and conclusions}
\label{con}
Our study has revealed a previously unknown streaming mechanism by internal waves at the lateral walls. It was only due to the absence of known streaming mechanisms in this experimental set-up that the previously unrecognized streaming mechanism was discovered. As such, it seems plausible that the mechanism we describe also played a role in other internal wave experiments. The horizontal particle advection observed by \cite{Ha10} in the vicinity of an internal wave attractor (see Fig. 7.3 of the thesis) appears to be such candidate, especially given the thin tank width of only $12$ cm. Strong evidence of the lateral-wall streaming at play is also provided by the three-dimensional numerical simulations by \cite{Br16}, where they found a lateral boundary-layer intensified mean circulation in the vicinity of wave attractor branches (their figures 7 and 8).
While it remains challenging to disentangle mean flow generation mechanisms in more complicated set-ups, we do believe that our lateral-wall streaming mechanism contributed to the induced mean flow observed both numerically and experimentally by \cite{Ki10} for tidal flow over three-dimensional topography. 
\\
\indent 
It should be noted that the presented laboratory experiments were not designed to investigate the lateral-wall streaming mechanism and that the theoretical background was only provided when the experiment was not available anymore. For future experiments we propose to include horizontal Particle Image Velocimetry measurements, to get a more direct observation of the induced mean flow, especially near the lateral walls. Additionally, it was unclear until detailed comparison with our theory that the particle sizes varied strongly among the eleven experiments. This prevented us from using the traditional expression \ref{wp} to determine the sinking velocity. 
Instead, we derived the sinking velocity accurately from the Doppler shifted column oscillation frequency, and subsequently used the expression \ref{wp} to infer the prevailing particle diameter, finding a factor $5$ difference among the eleven experiments. We believe that the large spread in particle size is caused by segregation of the carrier fluid in the newly designed particle injector. Changes in particles sizes among experiments may be prevented by using higher quality particles with more well-characterized particle sizes.
\\ 
\indent 
In the ocean, sites where internal waves propagate between two almost-vertical walls over distances longer than the channel-width are sparse. A topographic feature similar to our laboratory set-up is the $500$ m deep channel between two coral atolls studied by \cite{Ra18}. As the length of the channel, roughly $10$ km, exceeds the cross-channel width of $2$ - $3$, we may apply our simple expression \ref{baru_dim} for the return flow. Using the observed M2 velocity amplitude of $0.2$ m/s, a wave length of $800$ m (we estimated a quarter wave length from spatial variances in their figure 14), and \BV frequencies ranging from $10^{-3}$ to $0.5 \cdot 10^{-2}$, we find a (theoretical) induced mean return flow of $1$ - $5$ cm/s. This is small compared to the observed, bottom-intensified mean current in excess of $1$ m/s \citep{Ra18}. Nevertheless, it indicates that the streaming at the lateral walls can be relevant in the ocean. While it is legitimate to neglect the Coriolis effect 
at the coral atolls (taking it into account modifies the angle $\theta_{M2}$ by only 6\%), this may not be appropriate closer to the poles. Our analysis is not applicable to sites where the Coriolis frequency is similar to the strongest tidal component, typically M2. In spite of decreasing the slope $\tan^2 \theta = (\omega_0^2 - f_0^2)/(N_0^2 - \omega_0^2)$ upon incorporating the Coriolis effect $f_0$ - thereby reducing the strength of the lateral-wall streaming - we do believe that the overall streaming is intensified in the presence of rotation due to the appearance of additional non-linear terms associated with the rotational part of the wave field. 
\\
\indent 
The directly forced mean flow in the vicinity of the boundary may be relevant for the transport of suspended sediment, nutrients and litter that is (occasionally) lifted into the near-bottom water column. Our theoretical analysis has revealed that mean flow generation is strongest for wave frequencies near the \BV frequency (justifying the Coriolis effect to be neglected). Hence, we expect that small-scale internal wave packets associated with frequencies close to the local \BV frequency ($N_0$) may facilitate along-boundary particle transport upon oblique reflections at steep topography, `oblique' meaning that the incident and reflected beams do not fall into the same vertical slice. 
\\
\indent 
We speculate that there exists a continuous transition from our streaming mechanism at lateral (vertical) walls to the streaming mechanism over a flat bottom, recently investigated by \cite{RV18}. The viscous boundary layer description by \cite{KC95a, KC95b} for supercritical internal waves reflecting at inclined boundaries may be a good starting point to extend the lateral-wall streaming analysis to \emph{oblique} wave beam reflections at \emph{inclined} boundaries. Furthermore, we propose to compute the boundary streaming upon oblique reflection at inclined walls for wave \emph{packets}, both for well-studied spherical Gaussian-shapes \citep{Su10} as well as elongated beam-like packets \citep{FKA18}.


\section*{Acknowledgement}
E.H. and F.B. are  thankful to the organizers of the 2017 Les Houches summer school on Turbulent Flows in Climate Dynamics, where significant progress in writing this article was made. The collaboration was initiated at the FDSE summer school 2016 in Cambridge, UK, where D.M. introduced F.B. to the experimental results. F.B. is grateful for support by NWO Mathematics of Planet Earth grant 657.014.006. This work has been partially supported by the ONLITUR grant (ANR-2011-BS04-006-01) and partially achieved thanks to the resources of PSMN (P\^{o}le Scientifique de Mod\'{e}lisation Num\'{e}rique) of the ENS Lyon.


\section{Appendix}
\subsection{Misconceptions concerning Stokes Drift}
\label{stokes_drift}
\cite{Ho15} suggested that the Stokes Drift of the wave beam facilitated the observed particle transport, motivated by simple calculations adopted by \cite{Ha10}. This appendix points towards subtle misconceptions that led both \cite{Ha10} and \cite{Ho15} to misleading interpretations of their experimental results. \\
  By definition, the Stokes Drift is the difference of the time-averaged Lagrangian and Eulerian flow \citep{LH53}:
  \begin{equation}
   {\bf u}_S \ = \ < {\bf u}_L > \ - \ <{\bf u}> \ = \ \epsilon < \nabla {\bf u}({\bf x},t) \cdot \int_{0}^{t} {\bf u}({\bf x},t')dt'> \ + \ \mathcal{O}(\epsilon^2),
  \label{eq_stokes}
  \end{equation}
  with the time averaging over one wave period, $<\cdot> := 1/T \int_{t_0}^{t_0+T} \cdot dt$, starting at arbitrary time $t_0$. \cite{Ho15} made the inappropriate choice $t_0=0$ in his analysis, which corresponds to assuming all particles to enter the wave beam at one particular point in time, rather than at any (random) point in time during one wave cycle. Repeating the analysis by \cite{Ho15}, we find a particle column widening  without net horizontal displacement upon averaging $t_0$ over $[0,T]$. Both \cite{Ha10} and \cite{Ho15} assumed $<{\bf u}>=0$, an assumption that is not true at $\mathcal{O}(\epsilon)$, the leading order of the Stokes Drift. It is well known  \citep{Wu71,OM86,ZD15, Be18b} and straight forward to determine from Eq. \ref{eq_stokes} that the vertical Stokes Drift component is identical (with opposite sign) to the vertical induced mean velocity, $\bar{w}$, given by (\ref{barw}), such that the Lagrangian mean flow in the vertical direction vanishes at its leading order:
 \begin{equation}
 < w_L> \ =\  < w > +\  w_S \ = \ 0 \ + \ \mathcal{O}(\epsilon^2).
 \end{equation}
For two-dimensional fluids, mass conservation and the presence of vertical walls also require the horizontal Lagrangian mean flow component to vanish. This simple analysis stresses that horizontal cross-beam variations are necessary to generate net particle transport. 

\subsection{Numerical simulations}
\label{numerical_sim}
We solve (dimensionalized) Eq. (\ref{Psi}) for the stream function $\bar{\Psi}^{n}_0$ at time $t=n \ dt$ numerically on a rectangular domain, $(x,y)\in[0,80]\times [-8.5,8.5]$ cm$^2$, with $151$ grid-points in each direction, and time step $dt=1$ s. We employing standard central difference discretisation in space and Euler Backward (EB) in time:
  \begin{equation}
 \begin{split}
    \left( \tilde{L} - dt \ \nu \tilde{B} \right) \bar{\Psi}_0^{n+1} = \tilde{L} \bar{\Psi}_0^{n} +dt \ U_0^2 k_0 \bar{F}_y.
\label{EB}
 \end{split} 
 \end{equation}
 Here,  $\tilde{L}$ is the conventional 5-node discretised Laplace operator with Dirichlet boundary constraints, and $\tilde{B}$ is the conventional 13-node discretised Bi-harmonic operator, using Dirichlet boundary constraints for first-neighbour nodes outside the domain and Neumann boundary constraints for second neighbours. 
 While $\tilde{L}$ and $\tilde{B}$ are sparse, the related inverse, $\tilde{T} = ( \tilde{L} - dt \ \nu \tilde{B} )^{-1}$, computed only once and used iteratively to solve Eq. (\ref{EB}) forward in time, is an almost full $m \times m$-matrix, with $m=149^2$. Conveniently, most matrix elements of $\tilde{T}$ are negligibly small, and we can set matrix elements below a certain threshold (here $10^{-8}$) to zero - retaining only 20\% non-zero matrix elements - without noticeable effect on the simulation. 
 \\
 \indent
 Simulations with grid size $101 \times 101$ reveal increases of 1- 10\% in simulated particle displacements after $730$ s. This numerical error is acceptable, as it is of the same order as the approximations associated with the perturbation expansion and smaller than uncertainties of the experimental results. Simulations for the induced mean horizontal flow with free-slip boundaries predict particle displacements a factor $\sim10$ larger, clearly indicating that the induced mean flow is damped primarily by lateral wall friction. 

\subsection{Particle trajectory and sinking velocity from Doppler shift}
\label{particle_trajectory}
In order to compare the experimental results (\S \ref{obs}) on particle advection with the the theoretical wave-induced mean flow (\S \ref{theory}) we briefly discuss the relevant particle dynamics. 
\\
\indent
The time scale associated with adjustment to the motion of the fluid for neutrally buoyant spherical particles is $\frac{2 a^2 }{9 \nu }$ \citep{MR83}, where $a$ is the particle radius. For our experiments, the adjustment time for the suspended particles is of the order of a few milliseconds. As such, the particle motion, ${\bf u}_p$, can be described as a superposition of the mean particle sinking velocity, $\bar{w}_p$, and the motion of a fluid parcel at the particle's position, ${\bf x}_p(t) = [x_p(t), y_p(t), z_p(t)]$:
\begin{equation}
{\bf u}_p = {\bf u}({\bf x}_p(t),t) + {\bf \hat{z}} \bar{w}_p(z_p(t)).
\end{equation}
The Stokes particle sinking velocity is given by 
\begin{equation}
 \bar{w}_p(z) = \frac{2 a^2 g }{9  \nu} \left( 1 - z N_0^2/g - \rho_p/\rho_0 \right) <0, 
\label{wp}
\end{equation}
where $\rho_p = 1.055 \pm 0.01$ g/cm$^3$ is the particle density, which is larger than the fluid density at the bottom of the tank, $\rho_0 = 1.039 \pm 0.002$ g/cm$^3$. 
\\
\indent
As mentioned in \S \ref{setup_particles}, the particle radius $a$ varies strongly among the eleven experiments, making Eq. \ref{wp} useless to determine the sinking velocity. Instead, we make use of a Doppler shift to determine the particle sinking velocity, as explained in the following. We can  approximate the trajectory of a particle passing $z$ at time $t'$, by $z_p(t)= (t-t')\bar{w}_p$. 
Inserting this linearized particle trajectory into the leading order wave beam field quantity, e.g. $u_0(x,y,z,t) \propto \exp[i (k_x x + k_z z-\omega_0 t)]$, we find that  the velocity of the sinking particle column edge is given by $u_c(x,y,z,t) \propto \exp[ i(k_x + k_z (z-z_p(t)) - \omega_0 t]$, which oscillates at Doppler shifted frequency $\omega_c = k_z \bar{w}_p + \omega_0$. We can accurately extract the Doppler shifted column frequencies from the time series (see Fig. \ref{fig:f4}a,b), allowing us to determine the mean sinking velocity 
$$\bar{w}_p = \frac{\omega_c - \omega_0}{k_z} <0 $$
without knowledge of the particle sizes. Subsequently, we can determine the particle size from the sinking velocity, using Eq. \ref{wp}. For our experiments, we found that the particle diameter ranged from $40$ to $240$ $\mu$m, with an increase of particle size in consecutive experiments. The value of 200 $\mu$m, reported by \cite{Ho15}, falls into this range, but is clearly not representative for all experiments. As mentioned above, we believe that the large spread in particle sizes is caused by segregation of the carrier fluid that transports the particles from the particle reservoir to the particle injector.
 \\
 \indent
Note that due to the stratification, the density difference, $\bar{\rho}(z)-\rho_p$, reduces by approximately a factor $2/3$ from top to bottom in the laboratory experiments with $N_0=1.1 \pm 0.03$ rad/s, fresh water (with density $0.998$ g/cm$^3$) at the surface and particle density $\rho_p = 1.055\pm 0.01$ g/cm$^3$. For the linear stratification, it is straightforward to solve $\dot{z}_p^S(t)=\bar{w}_p^S = \bar{w}^S(z_p^S(t))$ for the associated particle trajectories starting at the particle injector, $z=h$, at time $t=t_0$: 
\begin{equation}
z_p(t)= h \left( -0.5 + 1.5 \exp\left[-2(t-t_0) \bar{w}_p(h)/(3h) \right]\right). 
\label{traject}
\end{equation}
This particle trajectory is superimposed on the space-time particle column displacement diagrams, Fig. \ref{fig:f3}.

\bibliographystyle{jfm}
\bibliography{manuscript}

\end{document}